\documentstyle[prl,aps,multicol,epsf]{revtex}
\tightenlines

\begin{document}

\title{Why high Tc is exciting.}

\author{ Jan Zaanen}
\address{Instituut-Lorentz for Theoretical Physics, Leiden University,
P.O. Box 9506, 2300 RA Leiden, The Netherlands}

\vspace{3mm}
\date{today}
\maketitle

\begin{abstract}
It is a common wisdom that the metallic state of solids has to do with a 
quantum-gas of particles which behave like non-interacting electrons. It has become clear during 
the last decade that systems of strongly interacting electrons are able to exhibit far more
interesting quantum-mechanical behaviors. The best evidence has been found in transition metal oxides,
especially so in the copper-oxide (high Tc) superconductors. Here I will present 
a sketch of the main developments. The plot is as follows: I am going to start off by shedding doubts on 
the established wisdom in metal physics (section I). In section II I will introduce the
`dynamical stripes', referring to an unprecedented form of quantum fluctuating order 
occurring on nanometer length- and picosecond time scales in the high Tc superconductors.
These dynamical stripes disappear at longer times where the physics of the superconductivity
emerges. This physics is highly  anomalous and I will discuss the popular notion
that it is not about quasiparticles but instead about the critical fluctuations associated 
with a quantum phase transition (section III). Such a phase transition should have to do with 
the disappearance of order but apparently this order cannot be detected by conventional
experiments (section IV). In the final section I will further illustrate this notion of `hidden order' with ideas
of our group in Leiden. This centers around the notion that a stripe phase carries a
very unusual form of order (`geometric order'), which can persist while the charge and
spin degrees of freedom of the stripe phase are quantum disordered, disappearing only
at the high dopings associated with the best superconductors.  
\end{abstract}

\begin{multicols}{2}
\narrowtext

\section{The unreasonable Fermi-liquid.}

Much of the present day electronics revolution would not have been possible
without the breakthroughs happening in the first half of the twentieth 
century in fundamental physics. In this era, the band-structure picture
of electrons in solids emerged. This picture is based on the notion that
the electrons behave approximately as non-interacting fermions and all
what remains to be done is to solve the Schr\"odinger equation describing 
the motion of a single electron through the potential exerted by the
static ion-lattice. The ultimate triumph of this idea was the explanation
of conventional superconductivity in terms of the Bardeen-Cooper-Schrieffer
(BCS) theory. According to this theory, superconductivity is a sibling of
the gas of non-interacting electrons. Under the influence of  any attractive force, these
fermions form pairs, and these pairs can subsequently
be viewed as a gas of bosons which have to Bose-Einstein condense in the
superconducting state.

This `paradigm' has been extremely successful. Reading the older textbooks
on the subject one gets the impression that it explains everything. Also
in modern fields like mesoscopic- and nano physics it is taken as the omnipotent
physical law, to the extent that theories are judged right or wrong pending
their conformation to the principle. However, is it obvious?  All
one has to realize is that electrons carry around a unit of electrical charge,
and in typical solids electrons are at average an Angstrom or so apart. Hence,
a simple estimate shows that these electrons repel each other with an
energy of order of electron-volts. How can such huge interactions be completely neglected? 

In conventional systems, like copper wire, silicon chips and neutron stars, the answer 
to this question is well established: at the
densities of interest the Fermi-energy, being the measure of the zero-point
kinetic energy, is even larger than the Coulomb energy and under this
condition one can pretend that the electrons do not interact. However, this
argument only works when the interactions are long ranged while the potentials
set by the ion lattice are weak. In this regard, copper cum suis are 
special and many systems have been identified where this argument 
does not apply. In such cases one faces a problem of principle. Even when
quantum mechanics can be neglected, dense systems of strongly interacting
particles (classical fluids) are very hard to describe, and quantum mechanics
makes this much harder. Nevertheless, for a long time it appeared that
nature was nice for theoretical physicists. Although one had to give up on
a complete description, it appeared that one could get away with a Fermi-gas
description at large length- and time scales: the Fermi-liquid notion of
Landau.  

Landau devised this strictly phenomenological description in the late 1950's for $^3He$, 
the first bad player which was identified. A number of other examples followed (like the
heavy fermion systems) and in the late 1980's the  Fermi-liquid notion was 
implicitly or explicitly considered to be the universal truth. Apparently, the reasons
why Landau's ideas were initially considered as an act of immense intellectual courage were
forgotten.  Consider $^3He$; surely
its low temperature collective physics is the Fermi-gas,  although the effective
fermions are much heavier than real $^3He$ atoms. However, at the same time its short
distance physics as measured by  neutron scattering  is barely different from that of the classical $^3He$
fluid found at higher temperatures. This is disturbing since it is well understood that
such a classical van der Waals fluid is far from being a non-interacting gas. It is
much closer to a crystal with defects and all motions are highly concerted.  Hence, a 
miracle is happening in quantum $^3He$: at short distances it is like a van der
Waals fluid kept in a highly collective motion by quantum mechanics, changing at
large distances in an entity where the effective $^3He$ atoms seem to fly straight 
through each other, not noticing each others influence except than for the Pauli
principle. How can this happen? I learned from Bob Schrieffer that nobody has a
clue and that this problem was simply abandoned, out of despair. 

These believes got badly shaken with the arrival of high Tc superconductivity. Despite
its obvious failure in this context, the Fermi-liquid was defended with a religious zeal. In hindsight,
the 1990's can be characterized as the era where the condensed matter community was
forced to abandon its paradigm, in a painful process which is not dissimilar from the
sociological dynamics described by the philosopher Kuhn.

For whatever reason, society seems to become more and more susceptible to a phenomenon
called hype. A good example is the recent outrage around e-commerce, but the physics
community is susceptible as well. High Tc superconductivity
started out like this in 1987. In the aftermath of the 1957 discovery of the BCS theory, a concerted effort
was organized to increase the superconducting transition temperature by designing materials based on the
BCS understanding. After 20 years or so this got stuck at a cold 24 K. The discovery of Bednorz and M\"uller
of a $T_c = 34$K superconductor in a copperoxide triggered a hype which got serious by the discovery
of a truly high temperature (90 K) copperoxide superconductor by Paul Chu {\em et al.} in early 1987.
For the next couple of years high Tc raged like a wildfire through the physics community, with the
predictable outcome of a severe hangover when it became clear that the rewards associated
with commercial superconductivity would not materialize. 

As an unplanned side effect, these
copper oxides were investigated in an unprecedented detail ($\pm 10^5$ papers) while it stimulated
the refinement of a variety of experimental techniques, varying from crystal growth to photoemission
and neutron scattering. The main result of this effort is that it once and for all proved the Fermi-liquid
and the BCS theory to be wrong. Despite this huge amount of experimental information, high Tc
superconductivity is still a mystery and it has only become more mysterious in the course of time. 
Being a mystery is not necessarily a sufficient condition for a flourishing science pursuit.  
However, it is a widespread sentiment among the specialists to be utterly fascinated by the problem,
and this sentiment rests on the  perception that the experiments guide us into novel but very fertile 
areas of physics research. The bottom-line is that since a couple of years the hangover is over.

\section{Mesoscopic quantum dynamics: stripes.}

The face of physics is a function of scale. In condensed matter the shortest scale is the lattice
constant and the physics is that of electrons moving in their (quasi) atomic orbitals, often called 
`chemistry'. In the established paradigm (e.g., copper) it is envisaged that these chemistry electrons
smoothly cross-over into Fermi-liquid quasi-electrons: all that happens is that the Coulomb interactions
of the lattice scale disappear due to metallic screening. In cuprates and other correlated oxides this
is an entirely different story. The lattice scale physics is reasonably well understood: it is the physics
of the doped Mott-insulator. Nearly all transition metal salts are insulators and this is due to the
dominance of the atomic Coulomb interactions localizing the electrical charges of the electrons (the
Mott-insulating state). The spins of the electrons can still move freely and one typically finds 
quantum-antiferromagnets. High Tc superconductivity emerges when a CuO based Mott-insulator is doped.
The active units are two dimensional CuO layers, separated by highly ionic oxidic layers containing
uninteresting elements like $La$. By chemical substitutions in the latter one can add or remove electrons
from the CuO layers. This introduces charge carriers into the planes, which delocalize quantum-mechanically.
It is by now well understood that this delocalization, leading to the metallicity,
is a highly collective affair. These moving charges scramble the spin system and as a result the
antiferromagnet quantum-melts locally and the charge carrier is surrounded by a droplet of quantum spin liquid.

The above picture is appropriate for a single, isolated carrier but it changes drastically at the carrier
densities of relevance to the superconductor which is rather high (one out of every eight unit cells 
contains a hole, or so). A collectivity sets in of a new kind having no precedent elsewhere:
the electron stripes. Instead of moving independently, the charge carriers organize on lines, `rivers
of charge', separated by Mott-insulating and antiferromagnetic domains, and these lines themselves are
subjected to quantum meandering motions on the CuO planes. As it turns out, these stripes can be brought
to a standstill by a variety of tricks (the so-called LLT lattice deformations, external magnetic
fields), all of which involve the removal of a relative small amount of kinetic energy from the electron system.  
Under these circumstances static stripe phases are formed where these `rivers of charge', form a  regular
structure which can easily be studied by  conventional means. It turns out that these stripe phases are
ubiquitous in doped Mott-insulators: they have not only been found in cuprates but in all
other doped Mott-insulators which have been studied up to now, like the manganites and the nickelates.
In Leiden, Hans Brom and coworkers are investigating the ordering dynamics of the stripes using NMR and NQR, and these
studies reveal that much remains to be understood, even when stripes are solidifying.

One statement is conclusive:  static stripe phases are not great for superconductivity. 
In fact, when stripes become static the system
tends to be (quasi) insulating. In a general sense they can be looked at as a special kind of Wigner 
(=electron) crystal. However, it appears that in the superconductors stripes are still around but
now as quantum fluctuating textures with a physical reality on mesoscopic length (1-10 nanometer)
and {\em time} scales (picoseconds). A crude analogy exists with the confinement phenomenon of
quantum-chromo dynamics. Stripes start to form at a scale of a couple of lattice constants as relatively
mildly fluctuating entities. However, upon increasing the scale these quantum fluctuations become
more and more severe to get truly out of hand at picosecond time scales. This is like the process occurring
in the QCD vacuum where at short distances the right objects are gluons and quarks. The quark/gluon fluctuations
become more and more severe in going to large distances with the effect that a qualitative change in the
physics occurs at the confinement scale where nuclear physics emerges. In the cuprate context, when the
stripe quantum fluctuations get out of hand, the physics of the high Tc superconducting state emerges. 

How do we know? The above picture is intimately linked to progress in `big gun' condensed matter
experimentation: photoemission and inelastic neutron scattering. Until recently, no experimental
means were available to directly probe this dynamical regime of mesoscopic lengths and -times.
How to probe fluctuating nanometer scale textures on 
a picosecond time scale? {\em Static} phenomena on  nanometer length scales are  easily accessible
with `conventional' nano-technology. Laser technology offers access to short times but averages 
automatically over micrometer lengths. However, both photoemisson and inelastic neutron scattering have
in principle access to electronic textures on nanometer length scales, which are fluctuating on picosecond time scales.
Due to an impressive progress over the last years,
the data have become good enough to be conclusive about the stripy mesoscopics in the high Tc cuprates.
Both experiments reveal features which indicate that both the spin- and the
electron dynamics acquire a one dimensional character, consistent with the stripe picture on
the aforementioned time- and length scales. 

This theme is actually more general. I refer to a recent  piece by Laughlin
and coworkers (`the middle way'), where it is argued that more surprises should be hidden in this mesoscopic
dynamical regime which will only reveal themselves when the appropriate experimental
machinery is available. For instance, a central mystery in biology is why proteins act
as flawless machines, and this is obviously related again to hard-to-probe mesoscopic dynamics.
Surely, neither photoemission nor neutron scattering have to say much about this mystery,
and Laughlin {\em et al.} argue that it should be a highest priority for experimentalists 
to figure out new machines giving access to these scales.

\section{Competing orders}

It is definitely not so that with the dynamical stripes the problem of high Tc superconductivity 
is solved. Along the lines of the QCD analogy of the previous section, at the stripe `confinement'
scale, the face of physics changes drastically and the long wavelength physics of the superconductor
is yet a completely different story. This long wavelength regime is easily accessible by conventional
condensed matter experimentation and it is in this regime where the mystery is most manifest. Superficially,
it has features which resemble a BCS superconductor, and there was a period that the opinion was
widespread that the cuprate superconductors somehow rediscovered BCS physics at sufficiently low
temperatures and large scales (compare with the $^3He$ example). However, in hindsight it appears
that a variety of anomalies were worked under the rug, while other anomalies became manifest
with the improving experimentation.

There is no debate regarding the metallic state realized at temperatures above the superconducting transition: it is
a quantum state of matter which has not a single feature in common with the Fermi-liquid. The newest
data indicate that this state continues smoothly into the superconducting state and it is
therefore an appropriate starting point to discuss high Tc's anomalies. One does not have to dig
deep: the simple property resistivity makes the point as clear as anything else. The resistivity
in the normal state of the best high Tc superconductors behaves in an extremely simple fashion: it just 
increases linearly with temperature, from $T_c\sim 100$K up to the highest temperatures measured
(1200 K). This is one example of the regularity of the high Tc phenomenon as I mentioned earlier: 
such a simple behavior should have a simple and  elegant explanation. Viewed
from a Fermi-gas perspective it is utterly unreasonable. When the electrical current is carried
by quasiparticles it has to be that the resistivity is a more interesting function of temperature
than a straight line. The reason is that the dissipation mechanism of the quasiparticle current
has to change as function of temperature. At low temperature, there are only other quasiparticles around
and the resistivity should be proportional to $T^2$. At intermediate temperatures phonons take
over and a non-universal  behavior is expected while at high temperatures the inelastic mean free
path becomes of order of the lattice constant and the resistivity should become temperature
independent. Hence, quasiparticle currents cannot cause linear resistivities and something else
is carrying the current! What else can it be instead?  There is only a single idea around which
makes sense. It is rooted in a simple and general idea: the current is carried by the quantum-critical 
fluctuation associated with a quantum phase transition.  

To appreciate the meaning of this sentence, one should not be scared by quantum-field theory. 
Quantum-field theory projects an image that it is an incomprehensible, overly mathematical 
affair which is barely ever of consequence, at least outside the realms of high energy physics. This is
quite besides the truth. It is better be regarded as a collection of powerful principles and
concepts which appear as increasingly simple and beautiful when one gets used to the idea.
The problem is just that it is a relatively novel discipline which emerged in its present
incarnation in the 1970's, and it is still to be included in the physics teaching programs. 
Apparently, field theory  is becoming alive in the context of high Tc superconductivity,
but also in quantum Hall and quantum magnetism, and this is the real reason behind the
perception that substantial progress is made in quantum condensed matter physics.

Quantum-field theory is about the quantum mechanics of systems with an infinity of degrees
of freedom and such systems are governed by principles which are different from those of
the few particle problems getting exposure in the textbooks. Its modern incarnation rests on the
path-integral formalism: a quantum system in D space dimensions can be viewed as a 
statistical physics problem in D+1 dimensions, with some special effects like the (anti) periodicity
in the time direction, Berry phases, etcetera. The role of temperature in the statistical physics
problem is taken by the coupling constant, measuring the strength of the quantum fluctuations,
while physical temperature enters the quantum problem as the inverse length of the imaginary
time axis.  

Taking this seriously, the idea of quantum-criticality becomes exceedingly simple. Statistical physics
is about phase transitions between ordered states, breaking some symmetry spontaneously, and disordered
states where the symmetry is restored. At the phase transition these two {\em collective} states of
matter are competing and when the phase transition is continuous this competition looks the same
on all scales, up to some short distance cut-off.  This universe becomes self-similar, which in turn
implies that correlation functions become algebraic, decaying like $x^{-\eta}$. Imagine now an ordered 
system at zero temperature, where the strength of the quantum-fluctuations can be tuned from the outside.
At some point a transition will follow to a quantum-disordered state. According to the path integral
formalism one can think about this quantum phase transition as a classical phase transition in space-time.
When this is a continuous  transition the path-integral formalism implies that in space-time a self-similar state is
realized: the quantum critical state. 
Since everything is algebraic, also the real-time dynamics becomes algebraic and dynamical
responses behave typically like power laws $\sim 1 / E^{\alpha}$ (`cusps', `branch cuts'), where $E$ is the energy 
pumped in the system from the outside. This is in marked contrast with quasi-particle excitations which show up as sharp
spikes (`poles') in response functions. Quasiparticles correspond with lumps of energy localized at some point in 
space-time and at the the critical point this is not possible because the quasiparticle breaks the scale invariance.
Instead, what one has are excitations (`quantum critical fluctuations') which fill all of space-time.

During the last few years,
the quality of the data on the high Tc superconductors, coming from photoemission, neutron scattering and optical experiments have
improved dramatically. Although these experiments measure different properties, they all show that the low lying
excitations of the have a cusp like nature, at least in the best cuprate superconductors.
This in itself already gives a strong support that we are dealing with a quantum-critical
system. However, a cross-check is possible. The above discussion refers to 
data taken at very low temperatures. A next specialty
of the quantum critical state is that temperature plays a quite peculiar role. As already stated, in the path-integral
formalism temperature enters as the inverse length of the imaginary time axis and at zero temperature the time axis is
infinitely long. However, at finite temperatures the time axis has a finite length and, therefore, {\em temperature
breaks the scale invariance in space-time}! Hence, besides the short distance cut-off (which can be argued to be 
easily of order 1000 K in the cuprates) the only scale in the problem is temperature itself! Using general properties of
finite size scaling one can argue that  zero frequency  properties measured at a finite temperature should be just
linearly proportional to temperature. I set out to explain why the resistivity is behaving like this, 
to find a most natural and general rational for this behavior in terms of quantum criticality!
This is not all because many other properties, including those
at finite energies and temperatures, find a natural explanation within this framework.

\section{Hidden order.}

Phase transitions tend to happen at isolated points in control parameter space and the same applies to
quantum-phase transitions. The most important control parameter is the coupling constant, parametrizing
the strength of the quantum fluctuations, and there are good reason to believe that this coupling constant
is in turn controlled by the amount of doping in the high Tc cuprates. The perfect quantum-criticality as
described in the previous section is found at a particular doping which is in the close vicinity of the
doping density where $T_c$ is at maximum. Going away from this point, bumps and wiggles (e.g., `pseudo-gap') appear in
physical properties which are consistent with the notion that, although the system cannot make a choice
at short distances/short times/high temperatures,  it has made up its mind at large distances/long
times/low temperatures. The phase transition is about two competing states of matter which are differing
in symmetry and away from the phase transition one of the two states  wins the contest. What are these states?

So much is clear that the quantum criticality is found right in the middle of the superconducting regime, and
surely these two states cannot be distinguished on their capacity to spontaneously break the gauge symmetry.
Something else is disordering. On the high doping side there is a sense that things start to look more normal
(BCS-like). Although far from being a proven fact, the notion is popular that this state is at least
symmetry-wise indistinguishable from a conventional superconductor. Hence, on the underdoped side a form
of order has to be present which is alien to a Fermi-liquid type superconductor. Remarkably, it is at present
completely unclear what this order is. Although it reveals its presence indirectly through the quantum-criticality,
it is apparently impossible to see it directly. Experimentalists have tried hard and found nothing. For this
reason it is called the {\em hidden order}.

To unravel the nature of this hidden order is the holy grail of high Tc. Given that it disappears at the maximum
Tc, the belief is widespread that it also will tell us something about the origin of the superconductivity. A
number of ingenious theoretical proposals are around where this link is made quite explicit. These all belong
to the class of theories based on the idea of spin-charge separation. The basic idea is quite simple. It is
asserted that the electrons fall apart in particles which carry the charge of the electron (`holons') and
excitations which carry its spin (`spinons'). The electron is a fermion which cannot bose condense to form a
superconductor. However, when the spinon carries away the fermionic character of the electron the holon is a
charge e boson and these bosons can in principle condense at a high temperature. Although
it is well established that spin-charge separation happens all the time in one dimensional systems, it has
appeared to be very difficult to demonstrate that it can happen in higher dimensional systems like the cuprate
superconductors. There is a variety of  uncontrolled theories around, based on spin-charge
separation, which all have the structure of a QCD-like gauge theory. These have in common that besides
the superconductivity very unconventional forms of long range order can occur. A first example is the
$SU(2)$ gauge theory by Patrick Lee and coworkers, suggesting that the hidden order is a {\em flux phase}.
In a flux phase, spontaneous electrical currents are flowing around the plaquettes in the lattice, setting
up a pattern of magnetic moments. These moments are not easy to observe because they are very small. However,
they are in principle observable and experimentalists have looked hard without finding anything. A more
recent idea is the Ising (Z2) gauge theory for spin-charge separation by Matthew Fisher and coworkers.
They propose a transition where the topological character of the gauge-vacuum is changing, coming up
with the prediction that in the hidden order phase the system should remember that it contained Abrikosov
flux lines, even when it is made non-superconducting. In the mean time, experiments have been performed
to check this prediction, with a negative outcome.

Besides these ideas centered around the spin-charge separation idea, there are a number of
other proposals around like the `conventional' flux phases (Varma, Laughlin and coworkers), as well
as the ideas of Sachdev and coworkers regarding a possible symmetry change of the superconducting
state itself. Although the books are not closed on the subject, it appears that these all suffer
from the same problem as the gauge theories: if around, these kinds of order should have been observed in
the mean time. Let me finally turn to a suggestion from our Leiden group. I find that it should
be taken seriously, for two reasons: (a) it is firmly based on stuff we know is real, the stripes, (b)
it is sufficiently outrageous to have a chance to be even true.

\section{Stripes and geometric order.}

I have now arrived at the point where the circle can be closed. I started out the discussion of high
Tc superconductivity with the anomalous behaviors called dynamical stripes, to give it no
further mention in the discussion of the long-wavelength quantum critical behaviors. Could it be that
the stripes and the quantum criticality have to do with each other? This is not obvious.
Recalling the analogy with QCD,
the experiments indicate that the stripy stuff literally disappears at a scale which is to be considered as small
in the context of the quantum-critical behaviors seen in the optimally superconductors. Although 
stripes occur at distances which are quite large as compared to the lattice constant, stripy things
should be around on macroscopic scales to be of relevance to the hidden order and the quantum
criticality. Stripes can be forced to order, so that this requirement is fulfilled, but for this to
happen one has to pay the price that superconductivity disappears. 

These statement are based on the perception that stripe order coincides with charge order and 
spin order (antiferromagnetism). These orders are easy to observe. The meaning of the dynamical stripes
is that the charge- and spin are indeed observed, but that they appear to be deep in the
quantum disordered regime, far from the phase transition. At the same time, I also argued that
the low energy physics of the cuprates emerges from this `stripy ultraviolet' which is so 
strikingly different from the electron ultraviolet of simple metals. Is there something in this
`short' distance stripe physics which we have overlooked, which can survive up to the macroscopic scale?

I gave in fact an incomplete characterization of stripe order in the second section. Stripes are more
than just spin and charge order. They carry yet another form of order which is so unfamiliar that it only
got formulated mathematically last year, although the community at large has been staring at it since 1994. 
The crucial experimental observation is that the charge
stripes are at the same time {\em Ising} domain walls in the stripe antiferromagnet. Every time one passes
a charge stripe the spin ordering pattern changes from up-down-up-down to down-up-down-up. Why this happens
is actually quite well understood. Many theoretical calculations, including the ones which led to the
theoretical discovery of the stripes by Gunnarsson and myself in 1987, have reproduced this `anti-phase 
boundarieness'. This physics is not essential for the further discussion and I refer the reader to the relevant
literature. Viewed from a symmetry perspective, the anti-phase boundarieness
 at first appears as an absurdity. Domain walls are
topological defects associated with a discrete (Ising-like, $Z_2$) symmetry. The problem is that the spin system
is a Heisenberg spin system: the antiferromagnetic order parameter can as well point along the z-,
x- or  y-direction, or anywhere in between. In theorist's jargon this is called $O(3)$ symmetry and such
a symmetry only allows for topological defects called skyrmions, which are entirely different from domain
walls. Hence, calling stripes domain walls in the spin system is just a misnomer.

As it turns out, the same basic $Z_2$ structure is present in the exact (Bethe-ansatz) solutions for the one dimensional
systems. It turns out to be responsible for the spin-charge separation: the Ising domain wall becomes a point (or
`particle') in one dimension and it binds to the electron, thereby `eating'  the
spin of the electron, turning it into a holon. In two dimensions, domain `points' turn into domain `lines' and
after binding the electrons to these domain lines one obtains precisely the stripes. Stripes might be called
`spin-charge separation in two dimensions' or, semantically more correct, 
the Luttinger liquid might be called a `mildly fluctuating one dimensional stripe phase'.

I have still not answered the question: the stripe (or holon) is a domain wall in what? Much helped by the 
highly advanced theory for the one dimensional case we only recently figured out the answer. This stuff
is sublattice parity. 

Sublattice parity refers to a {\em geometrical} property of the space in which the spin system lives. Given that
spin system is antiferromagnetic,  there is a crucial difference between a bipartite and non-bipartite embedding
space. A bipartite lattice is one which can be subdivided into two sublattices (A and B) and a simple square
lattice as realized in the cuprates is a good example. This subdivision can be done 
in two ways: $\cdots - A - B - A - B \cdots$ or $\cdots - B - A - B - A - \cdots$, and this `sublattice parity'
is obviously an Ising degree of freedom. Given that the nearest-neighbor interactions are by far the strongest,
one can realize a neat antiferromagnet on such a bipartite lattice by just putting, say, up-spins on the A- and
down spins on the B-sublattice. If the bipartiteness is destroyed, the spin system gets frustrated because
one can no longer satisfy the requirement that all neighboring spins are anti-parallel. In fact, the only
property of the embedding space which matters for the quantum antiferromagnet is if the embedding space is
bipartite or not.

Strangely, the Bethe-Ansatz solutions for the one dimensional systems show unambiguously
that the {\em electron charge binds to flips in the sublattice parity}. This defines a new bipartite space
in which the spins move and this is behind the explanation of spin-charge separation. Applying this one dimensional
recipe to the two dimensional case one obtains the stripes. The antiphase-boundariness as seen in the experiments
reveals that sublattice parity is around as a `hidden variable'. In the static stripe phases this 
sublattice parity is ordered, because every time one crosses a charge stripe the antiferromagnet reverses its
direction, without any exception.

Therefore, in the ordered stripe phases sublattice parity order is far from hidden. However,
it is not hard to imagine that sublattice parity actually goes undercover. A priori, there is nothing against the
theoretical possibility that the charge- and spin degrees of freedom quantum disorder, long before
sublattice parity order gives up. Imagine that the stripes are still intact lines, maintaining their domain wall
character, while these lines themselves form a quantum fluid. Only if one would take snapshots one would be
able to see that there are two different types of domains. If one waits too long one only sees the average,
characterized by equal amounts of A-B-A-B and B-A-B-A sublattice parity. 
One would expect that eventually at sufficiently high doping even sublattice parity order
should disappear. For instance, assuming that the overdoped state of the cuprates is a conventional BCS superconductor,
it has to be that sublattice parity order has disappeared because this kind of order is alien to a conventional
superconductor. Sublattice parity order has to disappear at a quantum-phase transition. 
Our suggestion is that high Tc's hidden order is sublattice parity order, which is disappearing at the famous quantum
critical point.

This is the simple idea. However, as a pleasant circumstance we found out that it is like opening a Pandora's box
of interesting theoretical physics. The reason is that sublattice parity is a geometrical structure, comparable but
far simpler than the physical space-time of general relativity. At the quantum phase transition, where the sublattice parity
order is disappearing, it is not so much the spins or the charges which go critical, but it is instead the effective space-time
in which they are living which is undergoing the critical fluctuations. This suggests that it has something to do
with the notoriously difficult problem of quantum gravity. However, we are much helped by the rather simple nature 
of `sublattice gravity'. With little success, theorists have attempted to map gravity on a gauge theory.
We found out that this strategy does become very successful in the present context: the gauge theory of relevance turns out
to be Wegner's $Z_2$ gauge theory. This is a lucky circumstance, because this is one of the few gauge theories which
is completely understood. Among others, the phase transition where sublattice parity order disappears turns out
to correspond with the confinement transition of the gauge theory which is known to be a second order transition.

Did we solve the problem? It is too early to say, because we are still facing considerable theoretical difficulties
associated with the fermion signs -- we seem to have a reasonable understanding of the theory when the matter (spin-, charge-)
fields are bosonic but there are theoretical- as well as experimental (nodal states) reasons to believe that a sign structure is
around. More worrisome, when the stripes are dynamical sublattice parity order gets very well hidden indeed! It seems
fundamentally impossible to observe it directly and only its indirect influence on especially the spin system
is accessible by experimental means. We do in fact predict a new type of spin-like order (the quantum spin-nematic)
which can be nailed down using conventional experimental means but it is not said that this state is also realized.
It is a weird state and it cannot be excluded that it is lying somewhere on an experimentalists shelf, as suspected
machine problem. However, we are not optimistic in this regard and there is nothing in the theory saying that
this state has to be realized.

The bottom-line is that sublattice parity order is suffering from the problem that it is too well hidden. For the
time being we take the liberty to continue working on it, finding inspiration in Dirac's principle that when the
mathematics is beautiful nature will do it.    

\section{Further reading.}

1. M. Buchanan, `Mind the pseudogap', Nature {\bf 409}, 8 (2001).\\
2. R. B. Laughlin, D. Pines, J. Schmalian, B.P. Stojkovic and P. Woynes, `The middle way', Proc. Nat. Ac. Sc. USA, {\bf 97},
28 (2000); R. B. Laughlin and D. Pines, `The Theory of Everything', Proc. Nat. Ac. Sc. USA, {\bf 97}, 32 (2000).\\
3. J. Orenstein and A. J. Millis, `Advances in the physics of high-temperature superconductivity', Science {\bf 288}, 468 (2000). \\
4. J. Zaanen, `Superconductivity: self-organized one dimensionality', Science {\bf 286}, 251 (1999);
`High-temperature superconductivity: stripes defeat the Fermi-liquid', Nature {\bf 404}, 714 (2000). \\
5. V. J. Emery, S. A. Kivelson and J. M. Tranquada, `Stripe phases in high temperature superconductors',
Proc. Nat. Ac. Sc. USA, {\bf 96}, 8814 (1999). \\ 
6. S. Sachdev, `Quantum Phase Transitions', Cambridge U. Press, New York (1999). \\
7. J. Zaanen, O.Y. Osman, H. V. Kruis, Z. Nussinov and J. Tworzydlo, `The geometric order of stripes and Luttinger liquids',
cond-mat/0102103. \\ 

\end{multicols}

\end{document}